\documentclass[12pt,preprint]{aastex}

\usepackage{graphicx}
\usepackage{amsmath}

\shorttitle{Interpreting Black Hole QPOs}
\shortauthors{Schnittman}

\begin{document}

\title{Interpreting the High Frequency QPO Power Spectra of Accreting
Black Holes}
\author{Jeremy D. Schnittman}
\affil{Department of Physics, Massachusetts Institute of Technology}
\affil{77 Massachusetts Avenue, Cambridge, MA 02139}
\email{schnittm@mit.edu}

\begin{abstract}
In the context of a relativistic hot spot model, we investigate
different physical mechanisms to explain the behavior of
quasi-periodic oscillations (QPOs) from accreting black holes. The
locations and amplitudes of the QPO peaks are determined by the
ray-tracing calculations presented in \citet{schni04a}: the black hole
mass and angular momentum give the geodesic coordinate frequencies,
while the disk inclination and the hot spot size, shape, and overbrightness
give the amplitudes of the different peaks. In this paper additional
features are added to the existing model to explain the broadening of
the QPO peaks as well as the damping of higher frequency harmonics in
the power spectrum. We present a number of analytic results that
closely agree with more detailed numerical calculations. Four
primary pieces are developed: the addition of multiple hot spots with
random phases, a finite width in the distribution of geodesic orbits,
Poisson sampling of the detected photons, and the scattering of
photons from the hot spot through a corona of hot electrons around the
black hole. Finally, the complete model is used to fit the observed power
spectra of both type A and type B QPOs seen in XTE J1550-564, giving
confidence limits on each of the model parameters. 
\end{abstract}

\keywords{black hole physics -- accretion disks -- X-rays:binaries}

\section{INTRODUCTION}
One of the most exciting results from the Rossi X-Ray Timing Explorer (RXTE)
was the discovery of high frequency quasi-periodic oscillations
(HFQPOs) from
neutron star and black hole binaries \citep{stroh96,stroh01,lamb02}. For
black hole systems, these HFQPOs are observed repeatedly at more or less
constant frequencies, and in a few cases with integer ratios
\citep{remil02,homan04,remil04}. These discoveries give the
exciting prospect of
determining a black hole's mass and spin, as well as testing general
relativity in the strong-field regime. 

To understand these observations more quantitatively, we have developed a
ray-tracing code to model the X-ray light curve from a collection of ``hot
spots,'' small regions of excess emission moving on geodesic orbits
\citep{schni04a,schni04b}. This hot spot model is motivated by the
similarity between
the QPO frequencies and the black hole coordinate frequencies near the
inner-most stable circular orbit (ISCO)
\citep{stell98,stell99a}, as well as the suggestion of a resonance leading to
integer commensurabilities between these coordinate frequencies
\citep{abram01,abram03}. \citet{stell99a} investigated primarily the QPO
frequency pairs found in LMXBs with a neutron star accretor, but their
approach can be applied to black hole systems as well. 

The basic geodesic hot spot model is characterized by the black hole mass and
spin, the disk inclination angle, and the hot spot size, shape, and
overbrightness. Motivated by the 3:2 frequency commensurabilites observed in
QPOs from XTE J1550-564, GRO J1655-40, and H1743-322
\citep{remil02,homan04,remil04}, we pick a radius for the geodesic
orbits such that the coordinate frequencies $\nu_\phi$ and $\nu_r$
will have a 3:1 ratio. Thus the 3:2 commensurability is interpreted as the
fundamental orbital frequency $\nu_\phi$ and its beat mode with the
radial frequency at $\nu_\phi-\nu_r$. Together with the assumption
that the concurrent low frequency QPOs are due to Lense-Thirring
precession at this same radius \citep{stell99b}, the location of the
QPO peaks uniquely determines the black hole mass and spin. Relaxing
the low frequency QPO criterion leaves a one-dimensional degeneracy in
the mass-spin parameter space which can be broken by an independent
determination of the binary system's inclination and radial velocity
measurements of the low-mass companion star [e.\ g.,
\citet{orosz02,orosz04}].

Given the black hole mass, spin, inclination, and the radius of the geodesic
orbit, the parameters of the hot spot model (i.\ e.\ the hot spot 
size, shape, and overbrightness, and the orbital eccentricity) are determined
by fitting to the amplitudes of the peaks in the observed power
spectrum. However, the model as described so far produces a perfectly
periodic X-ray light curve as the single hot spot orbits the 
black hole indefinitely. Such a periodic light curve will give a power
spectrum composed solely of delta-function peaks \citep{schni04a},
unlike the broad features in the observations.

In this paper we introduce two simple physical models to account for
this broadening of the QPO peaks. The models are based on analytic results,
then tested and confirmed by comparison with the three-dimensional
ray-tracing calculations of multiple hot spots \citep{schni04b}. These 
results draw repeatedly from the convolution theorem of complex
Fourier analysis, an essential tool for relating the behavior of the physical
system in the time domain with the more easily measured power spectrum in the
frequency domain. We find the power spectrum can be accurately modeled by
a superposition of Lorentzian peaks, consistent with the standard
analysis of QPO data from neutron stars and black holes
\citep{nowak00,bello02}. Many of the methods and results presented
here are equally valid for other QPO models such as diskoseismology
\citep{wagon01}, vertically-integrated disk oscillations
\citep{rezzo03}, toroidal perturbations \citep{lee02,lee04}, and
magnetic resonances \citep{wang03}.

In Section 2, we derive the effect of summing the light curves from
multiple hot spots with random phases and different lifetimes to give
a Lorentz-broadened peak in the power spectrum. Section 3 shows how a
finite width in the radii of the geodesic orbits produces a
corresponding broadening of the QPO peaks. In Section 4 we
develop two simple models for detector statistics and photon
scattering, both of 
which affect other features of the power spectrum such as the
continuum noise and the damping of high frequency harmonics, but do
not contribute to the broadening of the QPO peaks. Finally, all the
pieces of the model are brought together in Section 5 and used to
interpret the power spectra from a number of observations of
XTE J1550-564. We conclude with a discussion of open questions for the
hot spot model as well as directions for future work. 

\section{PEAK BROADENING FROM RANDOM PHASES}
In this section we derive the shape of a QPO peak in Fourier space
broadened by
the summation of multiple periodic functions combined with random
phases. There are many different accepted conventions for discrete and
continuous Fourier transforms \citep{press97}, so we begin by defining the
forward- and reverse-transforms between the time and frequency domains ($t$
and $\nu$). For a Fourier pair $f(t)$ and $F(\nu)$,
\begin{equation}
F_j = \frac{1}{N_s} \sum_{k=0}^{N_s-1} f_k e^{-2\pi i j k/N_s} \to
F(\nu) = \frac{1}{T_f}\int_0^{T_f} f(t) e^{-2\pi i \nu t} dt
\end{equation}
and
\begin{equation}
f_k = \sum_{j=0}^{N_s-1} F_j e^{2\pi i j k/N_s} \to
f(t) = T_f\int_{-\nu_N}^{\nu_N} F(\nu) e^{2\pi i \nu t} d\nu,
\end{equation}
where $f(t)$ is defined on the time interval $[0,T_f]$ and $\nu_N=1/(2\Delta
t)$ is the Nyquist frequency for a sampling rate $\Delta
t=T_f/N_s$. With this convention, $f(t)$ and $F(\nu)$ conveniently have the
same units and Parseval's theorem takes the form
\begin{equation}\label{spectral}
\int_0^{T_f} f^2(t)dt = T_f^2\int_{-\nu_N}^{\nu_N}F^2(\nu)d\nu.
\end{equation}
For such a time series $f(t)$, the power spectrum is defined as
$F^2(\nu)$, the squared amplitude of the Fourier transform.

Consider a purely sinusoidal function 
\begin{equation}
f(t) = A\sin(2\pi\nu_0 t+\phi),
\end{equation}
where $\phi$ is some constant phase. If
there are an integer number of complete oscillations within the time $T_f$ or
in the limit of $T_f\to \infty$, the Fourier transform of $f(t)$ will be 
\begin{equation}
F(\nu) = \left\{ \begin{array}{cl}
\frac{A}{2}e^{i(\phi-\pi/2)} & \nu = \nu_0 \\
\frac{A}{2}e^{-i(\phi-\pi/2)} & \nu = -\nu_0 \\
0 & {\rm otherwise} \end{array} \right. .
\end{equation}
If we then truncate the function $f(t)$ by multiplying it with a boxcar
window function $w(t)$ of length $\Delta T$, the convolution theorem gives
the transform of the resulting function $g(t)$:
\begin{eqnarray}
g(t) = f(t)w(t) \Leftrightarrow G(\nu)=(F\star W)(\nu).
\end{eqnarray} 
In the case where the window function is longer than a single period and
short compared to the total sampling time ($1/\nu_0 < \Delta T \ll T_f$), the
convolved power $G^2(\nu)$ can be approximated by 
\begin{equation}\label{sin1}
G^2(\nu) \approx \frac{A^2}{4T_f^2}
\frac{\sin^2[\pi(\nu \pm \nu_0)\Delta T]}{\pi^2(\nu \pm \nu_0)^2}.
\end{equation}

For $f(t)$ real, $F(-\nu)=F^\ast(\nu)$ and since we are primarily concerned
with the power spectrum $F^2(-\nu)=F^2(\nu)$, we will consider only positive
frequencies in the analysis below (unless explicitly stated otherwise).
Of course, when calculating the actual observable power in a
signal, both positive and negative frequencies must be included.

All the information about the phase $\phi$ of $f(t)$ and
the location in time of the window function is contained in the complex
phase of the function $G(\nu)$. This phase information is important
when considering the total power contributed by a collection of
signals, each with a different time window and random phase. 
When summing a series of complex functions with random
phase, the total amplitude adds in quadrature as in a two-dimensional
random walk. Therefore combining $N$ different segments of $f(t)$, each of
length $\Delta T$ and random $\phi$, gives a Fourier transform with amplitude
$\sqrt{N}|G(\nu)|$, and thus the net power spectrum is $NG^2(\nu)$.

The result in equation (\ref{sin1}) is valid only if every segment of
$f(t)$ has the exact same sampling length $\Delta T$ and frequency
$\nu_0$. This is not 
necessarily the case in the context of the hot spot model, in which
small regions of excess emission move along
geodesic orbits, modulating the flux periodically as photons
are relativistically beamed toward the observer. 
Three-dimensional MHD simulations of
accretion disks suggest that in general such hot spots are continually
being formed and destroyed with random phases, with a range of
lifetimes and orbital frequencies \citep{hawle01,devil03}.

For now, we consider the contribution from identical hot
spots, assuming that each one forms around the same radius with
similar size and overbrightness and survives for some finite time
before being destroyed. Over this lifetime, the hot spot produces a
coherent periodic light curve sampled by a window function of
duration $T$. Analogous to radioactive decay processes, we assume that
during each time step $dt$, the probability of the hot spot dissolving
is $dt/T_l$, where $T_l$ is the characteristic lifetime of the hot
spots. In this case, the differential probability distribution of
lifetimes $T$ for coherent segments is
\begin{equation}
P(T)dT = \frac{dT}{T_l}e^{-T/T_l}.
\end{equation}
Over a sample time $T_f \gg T_l$, if there are an average of $N_{\rm
spot}$ hot spots in existence at any time, the number of hot spots
formed with a lifetime between $T$ and $T+dT$ is given by 
\begin{equation}\label{distribution}
dN(T) = N_{\rm spot} \frac{T_f}{T_l^2}e^{-T/T_l}dT.
\end{equation}

Assuming for the time being that each coherent section of the light curve is
given by the sinusoidal $f(t)$ used above, we can sum all the individual
segments to give the total light curve $I(t)$ with corresponding power
spectrum
\begin{eqnarray}\label{lorentz1}
\tilde{I}^2(\nu)&=&\int_0^\infty G^2(\nu,T) dN(T) \nonumber\\
&=& N_{\rm spot}\left(\frac{A}{2\pi T_f}\right)^2 \int_0^\infty
\frac{\sin^2[\pi(\nu-\nu_0)T]}{\pi^2(\nu-\nu_0)^2}
\frac{T_f}{T_l^2}e^{-T/T_l} dT \nonumber\\
&=& 2N_{\rm spot}A^2\frac{T_l}{T_f} \frac{1}{1+4\pi^2 T_l^2 (\nu-\nu_0)^2}.
\end{eqnarray}
Hence we find the shape of the resulting spectrum is a Lorentzian
peaked around $\nu_0$ with characteristic width 
\begin{equation}\label{dec_width}
\Delta \nu = \frac{1}{2\pi T_l}.
\end{equation}

If this model is a qualitatively accurate description of how hot spots
form and dissolve in the disk, one immediate conclusion is that the
oscillator quality factor $Q$ can be fairly high even for relatively short
coherence times:
\begin{equation}
Q\equiv\frac{\nu_0}{\rm{FWHM}}=\pi T_l\nu_0.
\end{equation}
For example, if every hot spot has a lifetime of exactly four orbits
$(T_l=4/\nu_0)$, the central peak of the power spectrum 
$G^2(\nu,T_l)$ has coherence $Q=4.5$, about what one would expect from a
first-order estimate. However, after integrating over the exponential
lifetime distribution to get the Lorentzian profile of equation
(\ref{lorentz1}), the resulting quality factor is $Q=12.6$,
roughly a factor of
three higher. \citet{remil02} observe quality values of $Q\sim 5-10$
for the HFQPOs seen in XTE J1550-564, corresponding to typical hot
spot lifetimes of only 2-3 orbits.

In addition to the boxcar window, we have also tried other sampling
functions to
model the lifetime and evolution of the hot spots. Since the boxcar
window represents an instantaneous formation and subsequent destruction
mechanism, the resulting power spectrum contains significant power
at high frequencies, a general property of discontinuous functions. 

A smoother, Gaussian window function in time gives a Gaussian profile
in frequency space: 
\begin{equation}\label{gauss}
w(t)=\exp\left(\frac{-t^2}{2T^2}\right) \Leftrightarrow 
W(\nu) = \sqrt{2\pi}\frac{T}{T_f}
\exp\left(\frac{-\nu^2}{2\Delta \nu^2}\right)
\end{equation}
where again the characteristic width is given by $\Delta \nu = 1/(2\pi
T)$. After integrating over the same distribution of lifetimes
$dN(T)$ as above, we get the power spectrum
\begin{equation}\label{erfc}
\tilde{I}^2(\nu) = 4\pi N_{\rm spot}A^2\frac{T_l}{T_f}z^3
\left[\sqrt{\pi}(1+2z^2){\rm erfc}(z)e^{z^2}-2z \right],
\end{equation}
where we have defined 
\begin{equation}
z \equiv \frac{1}{4\pi T_l(\nu-\nu_0)}.
\end{equation}
For large $z$ (near the peak at $\nu=\nu_0$), equation (\ref{erfc})
can be approximated by the narrow Lorentzian
\begin{equation}
\tilde{I}^2(\nu) \approx 4\pi N_{\rm spot}A^2\frac{T_l}{T_f}
\frac{1}{1+48\pi^2 T_l^2(\nu-\nu_0)^2}.
\end{equation}

As with the boxcar window, the exponential lifetime distribution has the
effect of narrowing the peak of the net power spectrum compared with
that of a single segment of the light 
curve with length $T_l$. This smaller width can be
understood by considering the distribution of hot spot lifetimes
[eqn.\ (\ref{distribution})] and their relative contribution to the
total power spectrum [eqns.\ (\ref{sin1}, \ref{gauss})]. While there are
actually more segments with individual lifetimes shorter than $T_l$,
the few long-lived
segments of the light curves add significantly more weight to the
QPO peaks since $W(\nu=0,T) \propto T$ while $\Delta \nu \propto
1/T$. 

In fact, for any set of self-similar sampling functions
$w(t,T)=w(t/T)$, the corresponding power spectra $W^2(\nu,T)$ can be
approximated near $\nu=0$ as a Lorentzian:
\begin{equation}
W^2(\nu,T) \approx \frac{T^2}{T_f^2}
\frac{1}{1+\beta^2T^2\nu^2},
\end{equation}
with $\beta$ a dimensionless constant over the set of $w(t,T)$. The
characteristic
width of $W^2(\nu,T)$ is thus defined as $1/(\beta T)$. Integrating
over the lifetime distribution $dN(T)$, the net power function is
given by 
\begin{equation}
\tilde{I}^2(\nu) \approx \tilde{I}^2(\nu_0)
\frac{1}{1+12\beta^2T_l^2(\nu-\nu_0)^2}.
\end{equation}
We see now that the general effect of an exponential distribution of
sampling lifetimes is to decrease the peak width, and thus increase
the coherency, by a factor of $\sqrt{12}\approx 3.5$. 

In addition to the boxcar and Gaussian windows, another physically
reasonable model for the hot spot evolution is that of a sharp rise
followed by an exponential decay, perhaps caused by magnetic
reconnection in the disk. In this case, the light curve would behave
like a damped harmonic oscillator, for which the power spectrum is
also given by a Lorentzian. Interestingly, the resulting QPO peak
width is exactly
the same, whether we use a collection of boxcar functions with an
exponential lifetime distribution, or if we use a set of exponential
sampling functions, each with the same decay time. In the discussion
below, we will assume a boxcar sampling function and an exponential
lifetime distribution, with its corresponding Lorentzian power
spectrum. This also facilitates a direct comparison with observations,
where the QPO data is often fit by a collection of Lorentzian peaks
\citep{nowak00,bello02}.  

Due to the linear properties of the Fourier transform, the above
analysis, while derived assuming a purely sinusoidal signal with a
single frequency $\nu_0$, can be applied equally well to any periodic light
curve with an arbitrary shape. If each coherent section of the light
curve is written as
\begin{equation}\label{segsum}
f(t) = \sum_j A_j \sin(2\pi \nu_j t + \phi_j),
\end{equation}
then the total power spectrum (integrating over a distribution of coherent
segments with random phase) is simply the sum of the Lorentz-broadened
peaks:
\begin{eqnarray}\label{multi_lor}
\tilde{I}^2(\nu) &=& \int_0^\infty |G(\nu,T)|^2 dN(T) \nonumber\\
&=& 2N_{\rm spot}A_j^2 \frac{T_l}{T_f}\sum_j 
\frac{1} {1+4\pi^2 T_l^2(\nu-\nu_j)^2}.
\end{eqnarray}
Note that every peak in the power spectrum $\tilde{I}^2(\nu)$ has the
same characteristic width $\Delta \nu=1/(2\pi T_l)$. 

The sum in equation (\ref{segsum}) can be generalized to a Fourier
integral so that (\ref{multi_lor}) becomes
the convolution of the segment power spectrum $F^2(\nu)$ with a
normalized Lorentzian $\mathcal{L}(0,\Delta \nu)$ centered on
$\nu=0$ with width $\Delta \nu$: 
\begin{equation}\label{conv_lor}
\tilde{I}^2(\nu) = \int_{-\infty}^\infty \frac{F^2(\nu')d\nu'}
{1+\left(\frac{\nu-\nu'}{\Delta \nu}\right)^2} = 
[F^2\star \mathcal{L}(0,\Delta \nu)](\nu).
\end{equation}

Now we can apply our results to the light curves as calculated by
the original ray-tracing code for a single geodesic hot spot. First, the  
X-ray signal over one complete period is calculated to give the
Fourier components $A_j$ in (\ref{segsum}). For geodesic orbits, the
power spectrum $F^2(\nu)$ is concentrated at integer combinations of
the black hole coordinate frequencies $\nu_\phi$, $\nu_r$, and
$\nu_\theta$. In \citet{schni04a} we showed how the Fourier amplitudes
depend on orbital inclination and eccentricity as well as hot spot shape and
overbrightness. Given these frequencies $\nu_j$, amplitudes $A_j$, and a
characteristic hot spot lifetime $T_l$, the integrated power spectrum follows
directly from equation (\ref{multi_lor}).

\begin{figure}[tp]
\caption{\label{plotone} Simulated power density spectrum (crosses) from a
ray-tracing calculation of many hot spots on geodesic orbits with random
phases and different
lifetimes, along with an analytic model (solid line) of that power
spectrum. The black hole has mass $M=10M_\odot$ and spin $a/M=0.5$, giving
$\nu_r= 95$ Hz and $\nu_\phi = 285$ Hz. The hot spot orbit has an
eccentricity of 0.1 around a radius of $r_0 = 4.89M$ and an inclination of
$70^\circ$. The peaks have Lorentzian profiles with $\Delta \nu \approx
11$ Hz, corresponding to a characteristic hot spot lifetime of four
orbits.}
\begin{center}
\scalebox{0.8}{\includegraphics{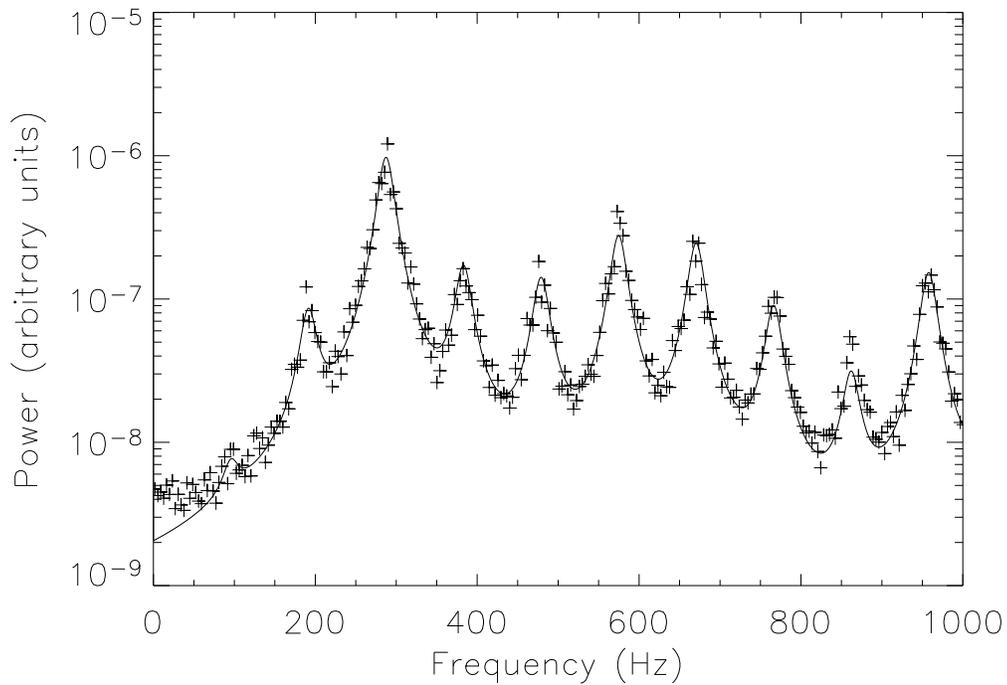}}
\end{center}
\end{figure}

Using the same ray-tracing code, we can also directly simulate the
light curve and corresponding power spectrum produced by many
hot spots orbiting with random phases, continually formed and
destroyed over each time step with probability $dt/T_l$
\citep{schni04b}. The power spectrum of such a simulation is shown in
Figure \ref{plotone} (crosses),
along with the analytic model (solid curve). We should stress
that this curve is \textit{not a fit to the simulated data}, but an
independent result calculated using the model described above. For this
particular example, the black hole has mass $M=10M_\odot$ and
spin $a/M=0.5$ with a disk inclination of
$i=70^\circ$. Each hot spot is on an orbit with $\nu_\phi=285$ Hz,
$\nu_r = 95$ Hz, and a moderate eccentricity of $e = 0.1$. This
particular orbit was chosen because of the 3:1 ratio in coordinate
frequencies, giving the strongest power at the modes
$\nu_\phi-\nu_r:\nu_\phi:\nu_\phi+\nu_r = 2:3:4$. Similar frequencies
appear to be the dominant peaks in the type A QPOs observed in XTE
J1550-564 \citep{remil02}.

The defining characteristic of QPO peaks broadened by the summation of
random phases is the uniform width of the individual peaks. For a
power spectrum with multiple harmonics and beat modes, each peak is
broadened by exactly the same amount, determined by the average lifetime of
the individual hot spots. Thus if we can measure the widths of multiple QPO
peaks in the data, the hot spot lifetime can be determined redundantly with a
high level of confidence.

\section{DISTRIBUTION OF COORDINATE FREQUENCIES}
In the previous section, we assumed a single radius for the hot spot
orbits. This assures identical geodesic coordinate frequencies for
different hot spots with different phases and lifetimes. However, this
assumption betrays one of the major weaknesses of the geodesic hot
spot model: there still does not exist a strong physical argument for
why these hot spots should form at one special radius or why that
radius should have coordinate frequencies with integer
commensurabilities. For now, we will be forced to leave that question
unanswered, but we can make progress by drawing on intuition gained
from other fields of
physics. If there does exist some physical resonance in the system that
favors these orbits, causing excess matter to ``pile up'' at certain
radii \citep{abram01,abram03}, then just like any other resonance, there
should be some finite 
width in phase space over which the resonant behavior is
important. The integer commensurability of the QPO peaks
suggests that closed orbits may be playing an important role in the
hot spot formation. If this is so, then some hot spots should also
form along orbits that \textit{almost} close, i.e.\ geodesics with
nearly commensurate coordinate frequencies. These orbits will have
guiding center radii similar to the critical radius $r_0$ for which
the geodesics form closed curves. 

Motivated by other processes in nature such as damped harmonic oscillators
and atomic transitions, we model the resonance strength as a function of
radius with a Lorentzian of characteristic width $\Delta r$. Then the
probability of a hot spot forming at a given radius is proportional to
the strength of the resonance there, giving a distribution of orbits
according to
\begin{equation}\label{lorentz_r}
P(r)dr = \frac{dr/(\pi\Delta r)}{1+\left(\frac{r-r_0}{\Delta r}\right)^2}.
\end{equation}
For a relatively small resonance width $\Delta r$, we can linearize the
coordinate frequencies $\nu_j(r)$ around $r=r_0$ with a simple Taylor
expansion: 
\begin{equation}\label{nu_taylor}
\nu_j(r) \approx \nu_{j0} + (r-r_0) \left. \frac{d\nu_j}{dr} \right|_{r_0},
\end{equation}
in which case the probability distribution in frequency space is also
a Lorentzian:
\begin{equation}\label{lorentz_nu}
P(\nu_j)d\nu_j = \frac{d\nu_j/(\pi\Delta\nu_j)}
{1+\left(\frac{\nu_j-\nu_{j0}}{\Delta \nu_j}\right)^2}.
\end{equation}
Here $\nu_j = \nu_\phi, \nu_\theta, \nu_r$ are the azimuthal, vertical,
and radial coordinate frequencies and $\nu_{j0}=\nu_j(r_0)$ are those
frequencies at the resonance center. 

\begin{figure}[tp]
\caption{\label{plottwo} Geodesic coordinate frequencies as a
function of radius for a black hole with mass $M=10M_\odot$ and spin
$a/M=0.5$. The radius of the inner-most stable circular orbit $r_{\rm
ISCO}$ is where $\nu_r \to 0$. The commensurate radius $r_0$ is where the
ratio of azimuthal to radial coordinate frequencies is 3:1.}
\begin{center}
\scalebox{0.8}{\includegraphics{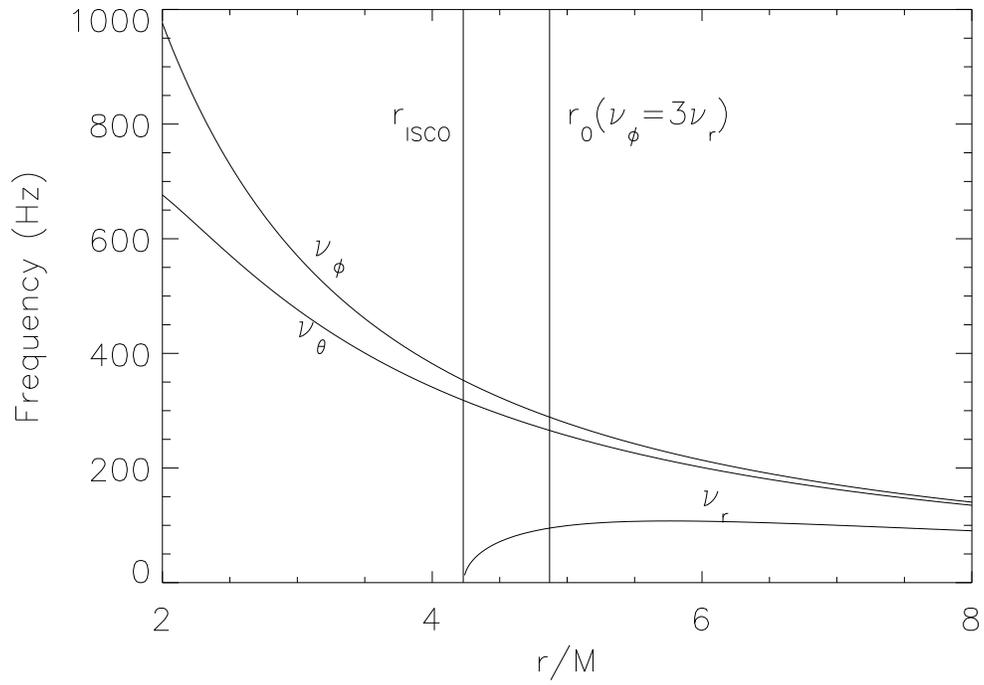}}
\end{center}
\end{figure}

For nearly circular orbits, the
coordinate frequencies (using geometrized units with $G=c=M=1$) are given by
\citet{merlo99} [following earlier work by \citet{barde72,perez97}]:
\begin{subequations}
\begin{equation}\label{nu_phi}
\nu_\phi = \frac{\pm 1}{2\pi(r^{3/2}\pm a)},
\end{equation}
\begin{equation}\label{nu_theta}
\nu_\theta = \nu_\phi \left[1\mp
\frac{4a}{r^{3/2}}+\frac{3a^2}{r^2}\right]^{1/2},
\end{equation}
and
\begin{equation}\label{nu_r}
\nu_r = \nu_\phi \left[1-\frac{6}{r}\pm \frac{8a}{r^{3/2}}
-\frac{3a^2}{r^2} \right]^{1/2},
\end{equation}
\end{subequations}
where the upper sign is taken for prograde orbits and the lower sign
is taken for retrograde orbits (the results below assume
prograde orbits, but the analysis for retrograde orbits is essentially
the same). These frequencies are plotted in Figure
\ref{plottwo} as a function of $r$ for a representative black hole
with mass $10M_\odot$ and spin $a/M=0.5$.
The radial frequency approaches zero at the ISCO, where geodesics can
orbit the black hole many times with steadily decreasing $r$. In the
limit of zero spin and large $r$, the coordinate frequencies reduce to
the degenerate Keplerian case with $\nu_\phi = \nu_\theta = \nu_r =
1/(2\pi r^{3/2})$. 

Generally, the power spectrum of the periodic light curve from a
single hot
spot orbiting at $r_0$ will be made up of delta functions located at
the harmonics of the fundamental $\nu_\phi$ and the beat modes with
$\nu_r$ and $\nu_\theta$. Considering for the moment only
planar orbits, the power will be concentrated at the frequencies
$\nu=n\nu_\phi\pm\nu_r$, where $n$ is some positive integer. In fact,
there will be additional peaks at $\nu=n\nu_\phi\pm 2\nu_r$ and even
higher beat-harmonic combinations, but for coordinate frequencies with
$\nu_\phi = 3\nu_r$, these higher modes are degenerate, e.\ g.\
$\nu_\phi+2\nu_r = 2\nu_\phi-\nu_r$. A careful treatment can
distinguish between these degenerate modes, but in practice we find
the power in the radial double- and triple-beats to be insignificant compared
to the single-beat modes at $n\nu_\phi\pm\nu_r$, so we limit our
analysis to these frequencies. 

From equations (\ref{nu_taylor}) and (\ref{lorentz_nu}), we see that
a QPO peak centered around $\nu=n\nu_\phi\pm\nu_r$ will be a
Lorentzian of width
\begin{equation}\label{broaden1}
\Delta \nu = \Delta r \left(n\frac{d\nu_\phi}{dr}\pm
\frac{d\nu_r}{dr}\right)_{r_0}.
\end{equation}
Unlike in the previous section where the random phases gave a
single width for every QPO peak, now each peak in the power
spectrum will be broadened by a different but predictable amount. Note in
particular how the peaks at the higher harmonics with $n>1$
will be significantly broader (and thus lower in amplitude) than the
fundamental. Another important feature evident from Figure
\ref{plottwo} and equation (\ref{broaden1}) is that, due to the
opposite-signed slopes of $\nu_r(r)$ and $\nu_\phi(r)$ around $r_0$, the beat
mode at $\nu_\phi+\nu_r$ remains very narrow, while the peak at
$\nu_\phi-\nu_r$ is quite broad. These features should play a key role
in using the power spectrum as an observable in understanding the
behavior of geodesic hot spots.

As in Section 2, the first step in producing a simulated
power spectrum is to calculate the Fourier amplitude in each mode with
the full three-dimensional ray-tracing calculation of emission from a
single periodic hot spot at $r_0$. Again, the linear properties of the
problem allow us simply to sum a series of Lorentzians, each
with a different amplitude, width, and location (peak frequency), to
get the total power spectrum. The amplitudes $A_j$ are given by the
ray-tracing calculations, the locations $\nu_j$ from the geodesic
coordinate frequencies and their harmonics, and the widths
$\Delta\nu_j$ from equation (\ref{broaden1}). 

Since the QPO peak broadening is most likely caused by a combination
of factors including the hot spots' finite lifetimes as well as
their finite radial distribution, the simulated power spectrum should
incorporate both features in a single model. Now the computational
convenience of Lorentzian peak profiles is clearly evident, since the
net broadening is given by the convolution of both effects and
the convolution of two Lorentzians is a Lorentzian: 
\begin{equation}\label{conv_2lor}
[\mathcal{L}(\nu_1,\Delta \nu_1)\star \mathcal{L}(\nu_2,\Delta
\nu_2)](\nu) = \mathcal{L}(\nu_{\rm tot},\Delta \nu_{\rm tot})(\nu),
\end{equation} 
where the peak centers and widths simply add: $\nu_{\rm tot} = \nu_1 +
\nu_2$ and $\Delta \nu_{\rm tot} = \Delta \nu_1+\Delta \nu_2$. In the
case where one or both of the Lorentzians is \textit{not} normalized,
the amplitude of the convolved function is given as a function of the
individual peak amplitudes and widths:
\begin{equation}
A_{\rm tot} = \pi\frac{A_1 A_2 \Delta\nu_1 \Delta\nu_2}
{\Delta\nu_1 + \Delta\nu_2},
\end{equation}
where $A_1$ and $A_2$ are the peak amplitudes of the respective
Lorentzians [$A_j=1/(\pi\Delta\nu_j)$ corresponds to a
normalized function.]

\begin{figure}[tp]
\caption{\label{plotthree} Power density spectrum (crosses) from a
ray-tracing calculation of many hot spots on geodesic orbits with 
different radii $r$ and thus different coordinate frequencies, along
with an analytic model (line) of that power
spectrum. The black hole has mass $M=10M_\odot$ and spin $a/M=0.5$,
while the average hot spot orbit has an eccentricity of 0.1 around a radius
of $r_0 = 4.89M$, as in Figure \ref{plotone}. The peaks have Lorentzian
profiles with $\Delta \nu$ given by equations (\ref{dec_width}) and
(\ref{broaden1}) with $T_l=100$ ms and $\Delta r = 0.05M$.}
\begin{center}
\scalebox{0.8}{\includegraphics{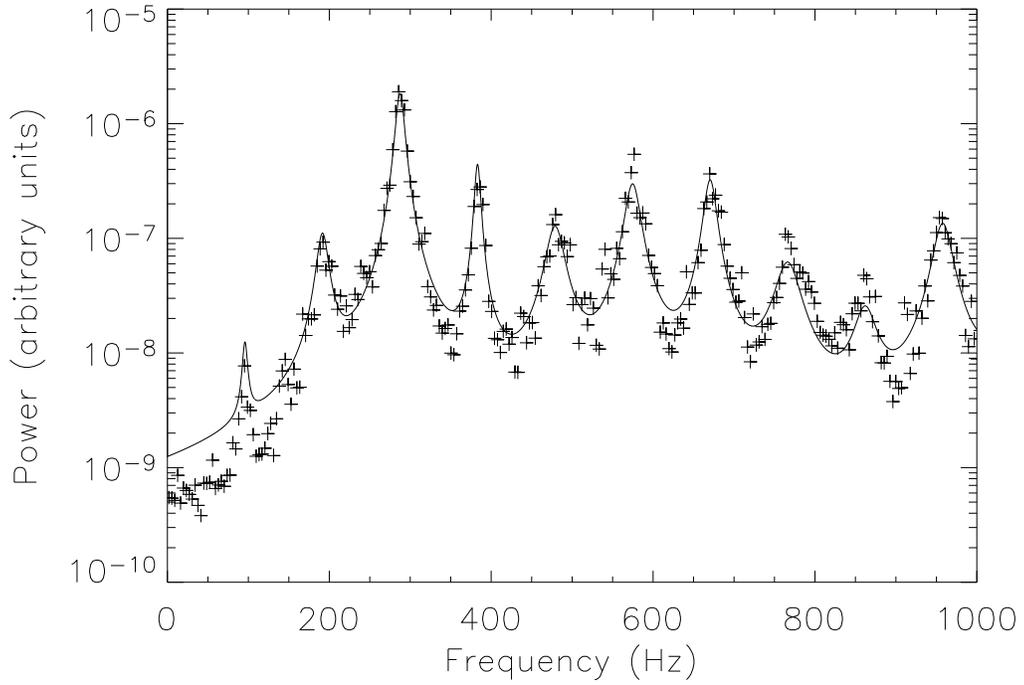}}
\end{center}
\end{figure}

Figure \ref{plotthree} shows the simulated power spectrum for a collection
of hot spots orbiting near the commensurate
radius $r_0=4.89M$ with a distribution width of $\Delta r =
0.05M$. All other black hole and orbital parameters are identical to
those in Figure \ref{plotone}. 
Both the random phase broadening described in Section 2 and the
effects of a finite resonance width are included in the model. Again, we
should stress that the solid line is not a fit to the data, but rather
an analytic model constructed from the sum of Lorentzian profiles as
described above. In this example, the hot spots have a typical
lifetime of 30 orbits, so the random phase
broadening contributes only $\Delta \nu \sim 1.5$ Hz for each
peak. This allows us to focus on the effect that a finite resonance
width has on the behavior of the QPO peaks at the coordinate
frequencies and their various beat harmonics. For a resonance width of
$\Delta r = 0.05M$, the peak widths due only to coordinate frequency
broadening are shown in Table \ref{tableone}.

The narrow peak at $\nu_\phi+\nu_r=380$ Hz and the neighboring broad
peak at $2\nu_\phi-\nu_r=475$ Hz are clearly visible in the simulated
data of Figure \ref{plotthree}. Precise measurements of each
peak's width may not come until a next generation X-ray timing mission, but
the qualitative behavior shown here should be detectable with the
current observational capabilities of RXTE. Combining equations
(\ref{dec_width}) and
(\ref{broaden1}) gives a system of equations that solve for the hot
spot lifetime $T_l$ and the resonance width $\Delta r$ as a function
of the QPO peak widths $\Delta \nu_j$. If we could accurately measure the
widths of only two peaks, both $T_l$ and $\Delta r$ could be
determined with reasonable significance. More peaks would give tighter
constraints and thus serve to either support or challenge the assumptions of
the hot spot model.

\section{PHOTON STATISTICS AND CORONAL SCATTERING}
Another feature that is important in modeling QPO power spectra
is based on the limitations of photon counting statistics by the RXTE
instruments. We can easily model this effect using the same ray-tracing
program employed above, now selecting only a small random fraction of
the rays to make up the light curve at each time
step. The resulting light curve is the ``standard'' light curve,
effectively multiplied by a random function in time with amplitude
described by a Poisson distribution, analogous to the sampling
functions used in Section 2. The corresponding Fourier transform is
the collection of original delta functions at the coordinate
frequencies, convolved with the transform of a Poisson-valued sampling
function.

For a Poisson distributed function $w(t)$ sampled over $N$ intervals
of $\Delta t$ with an average value of $\mu$, the corresponding Fourier
transform $W(\nu)$ is given by
\begin{equation}\label{poisson_transform}
W(\nu) = \left\{ \begin{array}{cl}
\mu & \nu = 0 \\
P(W) & \nu \ne 0 \end{array} \right. ,
\end{equation}
where for non-zero values of $\nu$, $W(\nu)$ is a complex random variable
with uniformly distributed phase and with amplitude given by
the Rayleigh distribution [e.\ g., \citet{groth75}]  
\begin{equation}
P(W)dW = \frac{2N}{\mu}We^{-W^2N/\mu}dW.
\end{equation}
The corresponding power $W^2(\nu\ne 0)$ appears as a background white noise
with mean 
\begin{equation}
\langle W^2 \rangle =\frac{\mu}{N}
\end{equation}
and variance
\begin{equation}
\sigma^2[W^2] = 4\frac{\mu^2}{N^2}
\end{equation}
constant over all frequencies. Unlike some of the
other instrumental contributions to the background power, the Poisson noise
is uncorrelated at different frequencies. 

As mentioned in the previous sections, the transform of a perfectly
periodic time series $f(t)$ will be zero everywhere except for a 
finite set of frequencies $\nu_j$ where $F(\nu_j)=A_j$. To compute the
simulated power spectrum of $f(t)$, multiplied by a Poisson sampling function
$w(t)$, we must first convolve the transform
functions $F(\nu)$ and $W(\nu)$ in frequency space and then take the
squared amplitude, giving the net power spectrum
$G^2(\nu)$, which will be composed of the
same delta functions that define $F^2(\nu)$, scaled by a factor of
$\mu^2$. In addition to these peaks, there will be a flat background
noise function $G^2(\nu \ne \nu_j)$ with the same type of probability
distribution as $W^2$. The net power spectrum is then given by
\begin{equation}\label{poisson_PDS}
G^2(\nu) = \left\{ \begin{array}{cl}
\mu^2 A_j^2 & \nu = \nu_j \\
P(G^2) & \nu \ne \nu_j \end{array} \right. ,
\end{equation}
where the noise spectrum $G^2(\nu \ne \nu_j)$ is again a random variable with
distribution function
\begin{equation}
P(G^2)dG^2 \propto \exp{\left(-\frac{N G^2}{\mu \sum A_j^2}\right)}
\end{equation}
and average power
\begin{equation}\label{avg_noise}
\langle G^2\rangle = \frac{\mu}{N}\sum_j A_j^2.
\end{equation}

One significant conclusion from this analysis is that the Poisson
sampling noise will not contribute to the broadening of the QPO
peaks. Since the Poisson noise is uncorrelated, a random sampling
window function combined with \textit{any} 
light curve [not just the strictly periodic $f(t)$ used here] will
maintain the Fourier properties of the original function $F(\nu)$
while adding a constant level of background noise characterized by
$P(W)$. Thus we can confidently apply our analysis from Sections 2 and
3 without worrying about additional broadening from random
emission and detection processes. Furthermore, these results give us
insights into an equally important part of the QPO power spectrum:
the background noise, which is critical when attempting to fit
Lorentz functions to the data and determining the amplitude of each
significant peak.

From equations (\ref{poisson_PDS}) and (\ref{avg_noise}), we see that
signal $G^2(\nu = \nu_j)$ to noise $G^2(\nu \ne \nu_j)$ scales as $\mu$,
thus giving a stronger detection for higher sampling rates, just as
expected. In the context of the hot spot model, we interpret $\mu$ to
be the average fraction of simulated light rays traced from the disk
to the detector, so typically $\mu \ll 1$. The actual value for
$\mu$ will depend on the spatial and time resolution of the
ray-tracing calculation as well as the luminosity and distance to the
source in question. For a source like XTE J1550-564 radiating with an
intensity of 1 Crab unit, the RXTE photon count rate in the range 2-60 keV is
$\sim 13,000$ counts s$^{-1}$ \citep{swank98}. With a typical QPO amplitude
of $1\%$ 
rms, this corresponds to roughly $10^{-2}-10^{0}$ photons \textit{from the
hot spot alone} per $50$ $\mu\rm{s}$ interval. 

If the background noise were due purely to Poisson statistics, sufficient
binning in the frequency domain would effectively remove the variance
of $G^2$ and allow us to subtract the mean background, leaving a pure
signal $G^2(\nu_j)$. However, for the proportional
counter array instrument on RXTE, there is an additional noise contribution
from the effect of detector ``deadtime'' that must be considered
during the data analysis process \citep{vande89,mitsu89}. This
additional correction is particularly important for high count
rates. For these high intensity sources, there are a number of
other detector contributions that must also be included. In practice, they
are generally combined and modeled as a single noise function
introduced in the data analysis pipeline, such as a
broken power law in frequency. 

Another simplified model we have included is that of scattering
photons from the hot spot off of a low-density
corona of hot electrons around the black hole and accretion disk. This
is known to be an important process for just about every observed
state of the black hole system
\citep{mccli04}. Unfortunately, it is also an extremely difficult
process to model accurately. Fortunately, for the problem of
calculating light curves and power spectra, a detailed description of
the scattering processes is probably not necessary. The most important
qualitative feature of the coronal scattering is a smearing of the hot
spot image: a relativistic emitter surrounded by a cloud of scattering
electrons will appear blurred, just like a lighthouse shining its beam
through dense fog. The effect is even more 
pronounced in the black hole case, where the hot spot orbital period
is of the same order as the light-crossing time of a small corona,
thus spreading out the X-ray signal in time as well as space. 

Since the
scattered photons are often boosted to higher energies, a coherent phase lag
in the light curves from different energy channels could be used to estimate
the overall scale length of the corona. \citet{vaugh97} have observed this
effect in neutron star QPOs and infer a scattering length of
$\lambda\sim 5-15M$ for an optical depth of $\tau \sim 5$ in the source 4U
1608-52. \citet{ford99} perform a similar analysis for black holes,
including the possibility for an inhomogeneous corona, and derive a much
larger upper limit $(\lambda \sim 10^3M)$ for the size of the
corona. In either case, the qualitative effect will be the same: the
damping of higher harmonic features in the power spectrum of the X-ray
light curve. 

The simple model we introduce is based on adding a random
time delay to each photon detected from the hot spot. The
distribution of this time delay is computed as follows: we
fix the optical depth to be unity for scattering though a
medium of constant electron density, so each photon is assumed to
scatter exactly once between the emitter and the observer, thus
determining the length scale of the corona as a function of
density. In this case, the probability of scattering after a distance
$l$ is
\begin{equation}
P(l)dl = \frac{dl}{\lambda}e^{-l/\lambda},
\end{equation}
where $\lambda$ is the photon mean free path in the corona. 

Next, due to the
likely existence of an optically thick disk around the black hole equator, we
assume that the photon scattering angle is less than $\pi/2$ (we
define the scattering angle $\theta'$ as the angle between the
incoming and outgoing wave vectors, so a straight path would correspond to
$\theta'=0$). In
other words, only photons emitted in a hemisphere facing the observer
can ultimately be scattered in the observer's direction. For a photon
emitted at an angle $\theta$ to the observer, scattering at a distance
$l$ from the source produces an additional photon path length of
$d=l(1-\cos\theta)$, assuming for simplicity a flat spacetime
geometry. While the photons are emitted with an isotropic distribution
in, the scattering distribution is \textit{not} isotropic. Since the
scattering geometry requires that $\theta=\theta'$, we only detect a
subset of the photons emitted with an angular distribution
in $\theta$ that satisfies this relationship.
In the limit of low-energy photons $(h\nu \ll m_ec^2)$ and elastic
scattering, the classical Thomson cross section $\sigma_T$ is used:
\begin{equation}
\frac{d\sigma}{d\theta'} =
\frac{3}{8}\sigma_T \sin\theta' (1+\cos^2\theta').
\end{equation}
Integrating this distribution over all forward-scattered photons
$(\theta' < \pi/2)$, we 
find the average additional path length to be $\langle d \rangle = 7l/16$. Since the time
delay is the path length divided by the speed of light $c$, scattering
once in the corona adds a time delay $\Delta t$ to each photon with
probability 
\begin{equation}
P(\Delta t)d(\Delta t) = \frac{d(\Delta t)}{T_{\rm scat}}
e^{-\Delta t/T_{\rm scat}},
\end{equation}
where the average scattering time is given by $T_{\rm scat} = 7\lambda/16c.$

Applied to the ray-tracing model, this has the effect of smoothing out
the light curve 
with a simple convolution in the time domain of the original signal
$f(t)$ and the time delay probability distribution function
$P(\Delta t)$. The Fourier transform of the resulting light curve is the
product of the two transforms $F(\nu)$ and $\tilde{P}(\nu)$, where
\begin{equation}
\tilde{P}(\nu) = \frac{1}{1+2\pi i T_{\rm scat}\nu}.
\end{equation}
When we square the product to get the
power spectrum $G^2(\nu) = F^2(\nu)\tilde{P}^2(\nu)$, the scaling
factor is yet again a Lorentzian: 
\begin{equation}\label{scatter_PDS}
G^2(\nu_j) = \frac{A_j^2}{1+(\nu_j/\Delta \nu_{\rm scat})^2},
\end{equation}
where the scale of frequency damping is given by 
\begin{equation}
\Delta\nu_{\rm scat} \equiv 1/(2\pi T_{\rm scat})
\end{equation}
and $A_j$ are the delta function amplitudes of $F(\nu)$ 
as defined above. This analytic result is perhaps a case where the
ends justify the means. Our model for electron scatting in the corona
is extraordinarily simplified, ignoring the important factors of
photon energy, non-isotropic emission, multiple scattering events in a
non-homogeneous medium, and all relativistic effects. However,
assuming that almost any analytic model would be equally (in)accurate,
at least the treatment we have applied proves to be computationally
very convenient.

\begin{figure}[tp]
\caption{\label{plotfour} Power density spectrum from a single hot
spot light curve where the emitted photons are scattered exactly once
each by a uniform corona of electrons. The simulated spectra
are plotted as dots and asterices, while the analytic model is a solid
line. In (a), the mean free path to scattering is $\lambda = 10M$,
while (b) represents a much larger, low density corona with $\lambda =
100M$.} 
\begin{center}
\scalebox{0.5}{\includegraphics{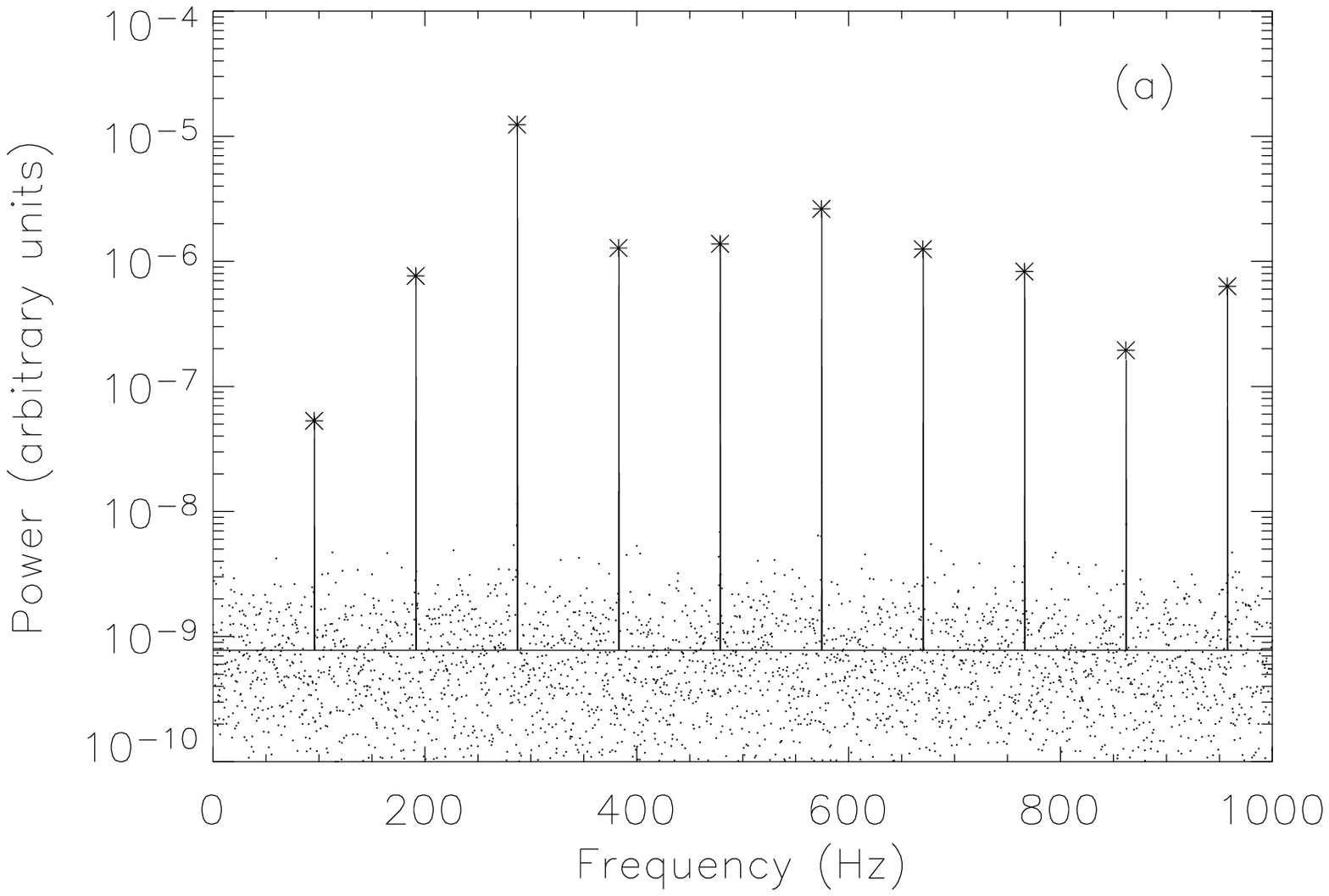}}
\scalebox{0.5}{\includegraphics{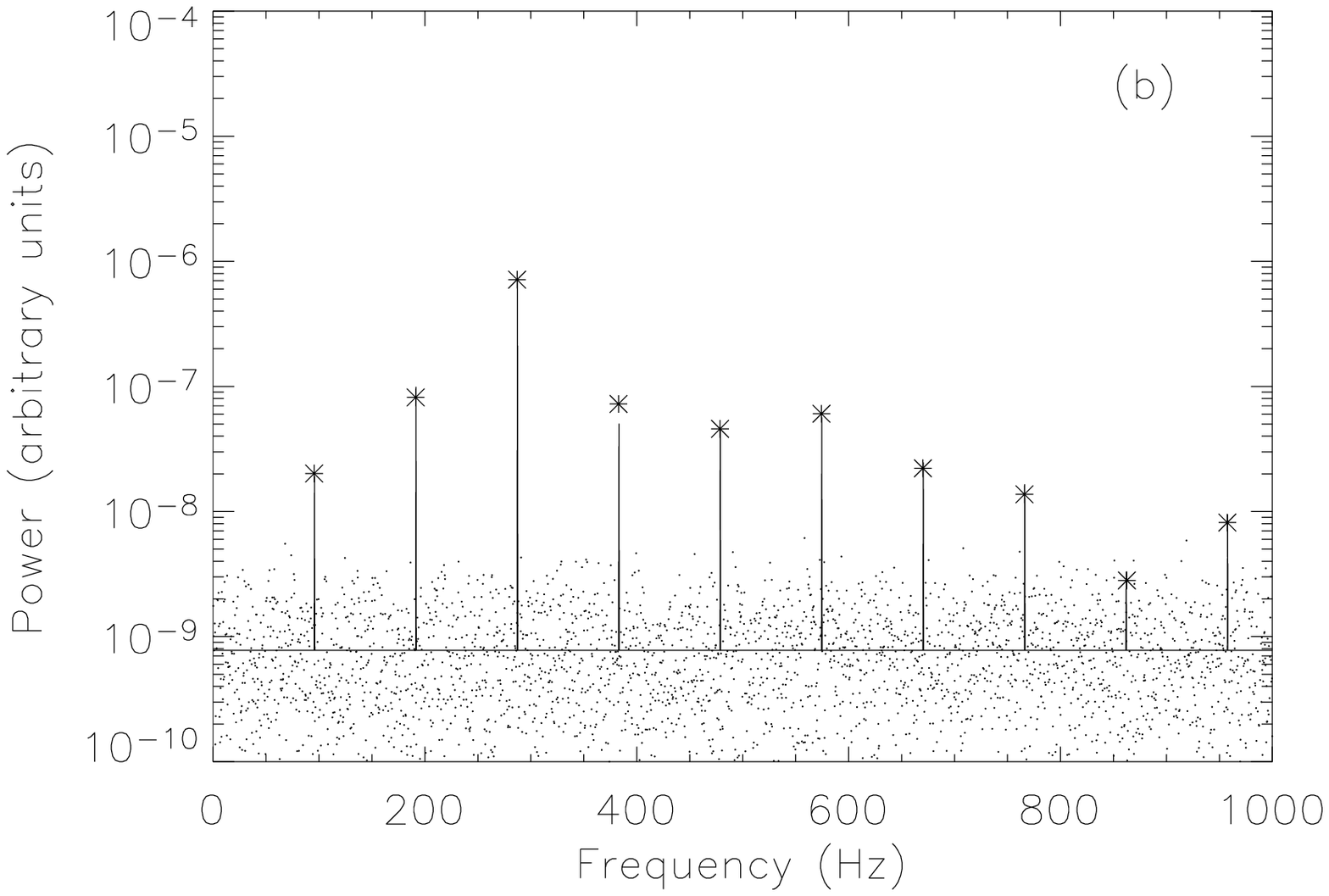}}
\end{center}
\end{figure}

Equation (\ref{scatter_PDS}) states that the
resulting power spectrum of the scattered light curve is a set of delta
functions, with the higher harmonics damped out by the effective
blurring of the hot spot beam propagating through the coronal
electrons. A simulated power spectrum is shown in Figure \ref{plotfour}a for a
scattering length of $\lambda = 10M$, comparable to the size of the
hot spot orbit. Figure \ref{plotfour}b shows the effect of a larger,
low-density corona with scale length $\lambda = 100M$, corresponding
to a longer convolution time and thus stronger harmonic damping. The
white background noise (Poisson noise with $\mu$ = 1) in both cases is
due to the statistics of the random scattering of each photon from one
time bin to another. The simulated spectra are plotted as dots
(asterices at $\nu_j$ to highlight the peaks) and the analytic model is a
solid line. 

As in the model for Poisson sampling, we see that the coronal
scattering should not contribute to the broadening of the QPO
peaks. However, it will have a very significant effect on the overall
harmonic structure of the power spectrum, particularly at higher
frequencies. \citet{schni04a} show a similar result caused by the
stretching of the geodesic blob into an arc along its path, also
damping out the power at higher harmonics. In this context, it is now
clear that the arc damping can be modeled analytically by
interpreting the stretching of the blob in space as a convolution of
the light curve in time. If the stretched hot spot
has a Gaussian distribution in azimuth with length $\Delta \phi$, the
original X-ray light curve will be convolved with a Gaussian window
of characteristic time $\Delta t = \Delta \phi/(2\pi\nu_\phi)$. Equation
(\ref{gauss}) gives the corresponding scaling factor $W(\nu)$ in the
frequency domain (replacing $T$ with $\Delta t/2$). The exponential
damping of $W(\nu)$ is stronger than the 
Lorentzian factor at higher frequencies, but both effects (coronal
scattering and hot spot stretching)
are probably important in explaining the lack of significant power
in the harmonics above $\sim 500$ Hz in the RXTE observations.

\section{FITTING QPO DATA FROM XTE J1550-564}
In this section we combine all the pieces of the model
developed above and apply the results to the RXTE data from type
A and type B QPOs observed in the low-mass X-ray binary XTE
J1550-564. To compare directly with the data from \citet{remil02}, we
need to change slightly our normalization of the power
spectrum. Following \citet{leahy83} and \citet{vande97}, we define the
power spectrum $Q(\nu)$ so that the total power integrated over
frequency gives the mean square of the discrete light
curve $I_j=I(t_j)$: 
\begin{equation}\label{Q_rms}
\int_{\nu>0}^{\nu_N} Q(\nu) d\nu =
\frac{1}{N_s}\sum_{j=0}^{N_s-1}\left(\frac{I_j-\langle I
\rangle}{\langle I \rangle}\right)^2, 
\end{equation}
where $I_j$ is sampled over $j=0,...,N_s-1$ with average value
$\langle I \rangle$. In terms of the power spectra used in Sections 2
and 3, $Q(\nu)$ is given by
\begin{eqnarray}
Q(0) &=& 2 \nonumber\\
Q(\nu) &=& 2 T_f\frac{\tilde{I}^2(\nu)}{\tilde{I}^2(0)},
\end{eqnarray}
which has units of $[(\rm{rms/mean})^2 \rm{Hz}^{-1}]$.

As we described in the introduction, the hot spot model is constructed
in a number of steps. These steps result in a first approximation for
the black hole and hot spot model parameters, after which a $\chi^2$
minimization is performed to give the best values for each data set.
\begin{itemize}
\item The black hole mass and the inclination of the disk are given by
optical radial velocity measurements \citep{orosz02}. We take
$M=10.5M_\odot$ and $i=72^\circ$ as fixed in this analysis. 
\item The black hole spin is determined by matching the
frequencies of the HFQPOs to the geodesic coordinate frequencies such
that $\nu_\phi=3\nu_r$ at the hot spot orbit, giving $a/M \approx 0.5$
for $\nu_\phi \approx 276$ Hz. The small uncertainties in the measured
value of $\nu_\phi$ can thus be interpreted indirectly as constraints on
the mass-spin relationship.
\item The orbital eccentricity and hot spot size and overbrightness
are chosen to match the total amplitude of the observed
fluctuations. We use a moderate eccentricity of $e=0.1$, but find the
peak amplitudes are not very sensitive to this parameter
\citep{schni04b}. The question of overbrightness is still an area of
much research, since the nature of the background disk is not well known
during the ``steep power law'' state that produces the HFQPOS
\citep{mccli04}. In practice, we set the hot spot emissivity constant
and then fit an additional steady-state background flux $I_B$ to the
variable light curve.
\item The hot spot arc length and the coronal
scattering time scale are chosen to fit the relative amplitudes of the
different QPO peaks. 
\item The hot spot lifetime and the width of the resonance around $r_0$
are chosen to fit the width of the QPO peaks.
\item The observed flux gives the detector count rate, which is added
to the white noise from the coronal scattering statistics. In practice, for
power spectra with sufficient binning statistics (corresponding to long
observations in time), the Poisson noise reduces to a flat spectrum which is
subtracted from the data before fitting the model parameters.
\item As a final step, we include an additional power law component
$\propto \nu^{-1}$ to account for the contribution
due to turbulence in the disk [e.\ g.\ \citet{mande99}] not accounted for by
the hot spot model. Instrumental effects such as the detector deadtime are
combined with the turbulent noise to give a simple two-component
background spectrum: 
\begin{equation}
Q_{\rm noise}(\nu) = Q_{\rm PL}\nu^{-1}+Q_{\rm flat}.
\end{equation}
\end{itemize}

After determining the fixed amplitudes $A_j$ with the ray-tracing
calculation, we minimize the $\chi^2$ fit over the following parameters:
orbital frequency $\nu_\phi$, hot spot lifetime $T_l$, resonance width
$\Delta r$, scattering length $\lambda$, hot spot arc length $\Delta 
\phi$, steady state flux $I_B$, and the background noise components $Q_{\rm
PL}$ and $Q_{\rm flat}$. The best fit parameters are shown in Table
\ref{tabletwo}, along with $1\sigma$ ($68\%$) confidence limits. These
confidence
limits are determined by setting $\Delta\chi^2 < 7.04$, corresponding to
six ``interesting'' parameters of the hot spot model, holding the noise
components constant \citep{avni76,press97}. We find that $Q_{\rm PL}$ and
$Q_{\rm flat}$ are almost identical for both data sets, supporting the 
presumption that they are indeed a background feature independent of the hot
spot model. 

\begin{figure}[tp]
\caption{\label{plotfive} Comparison of hot spot model power spectrum
(line) with data 
(crosses) from XTE J1550-564. (a) Type A QPO, dominated by a narrow peak at
$\nu_\phi \approx 280$ Hz. (b) Type B QPO, dominated by a broad peak at
$\nu_\phi-\nu_r \approx 180$ Hz. The best fit model parameters for each
data set are shown in Table \ref{tabletwo} and the resulting QPO
amplitudes and widths are shown in Table \ref{tablethree}.} 
\begin{center}
\scalebox{0.5}{\includegraphics{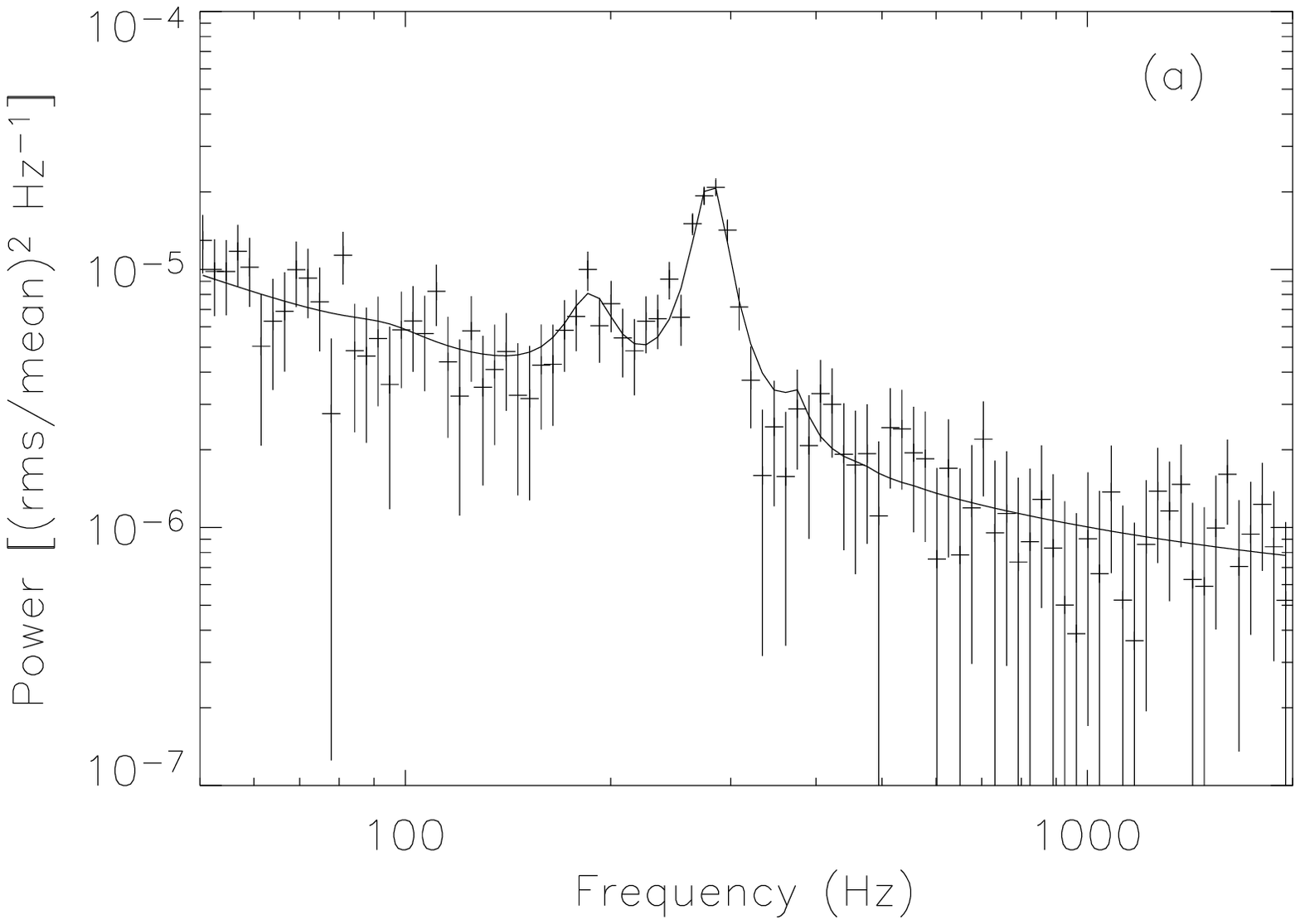}}
\scalebox{0.5}{\includegraphics{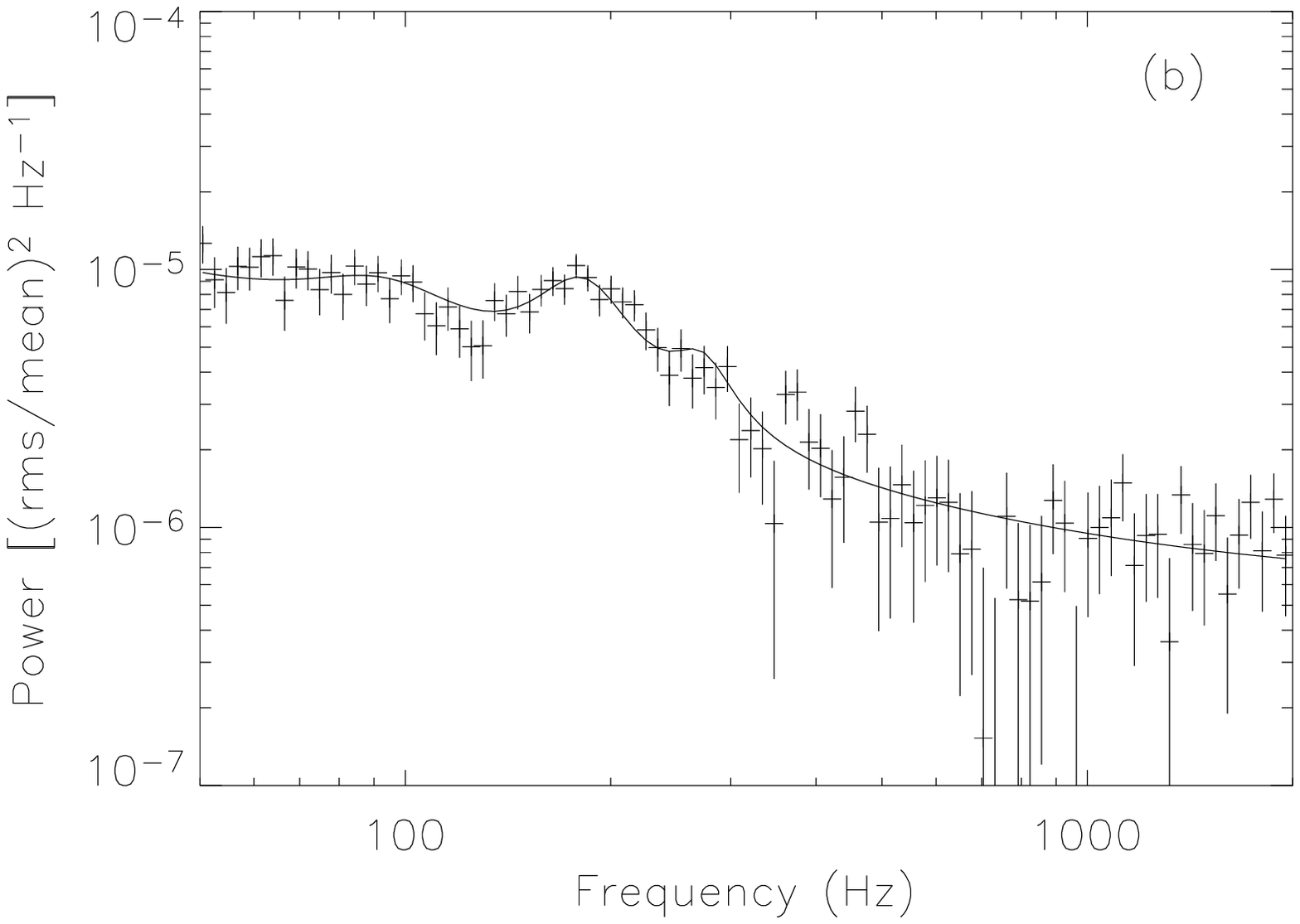}}
\end{center}
\end{figure}

In Figure \ref{plotfive} we show the
observed power spectra for type A and type B QPOs, as reported in
\citet{remil02}, along with our best fit models. The type A QPOs are
characterized by a strong, relatively narrow peak at $\nu_\phi \approx
280$ Hz, corresponding to $\nu_\phi$ in our model, with a minor peak
of comparable width at $\nu_\phi-\nu_r \approx 187$ Hz. Type B
QPOs on the other hand, have a strong, broad peak around 180 Hz with a
minor peak at 270 Hz. This implies a longer arc,
damping out the higher frequency modes, and a shorter average lifetime,
broadening the peaks. Both types of QPO suggest a very narrow
resonance width $\Delta r$, yet the current data does not constrain this
parameter very well. Thus we assume the majority of the peak broadening
is caused by the addition of multiple hot spots with random phases and a
characteristic lifetime of $\sim 3$ orbits for the type A QPOs and
about half that for type B.

We performed a covariance analysis of the parameter space near the $\chi^2$
minimum to identify the best-constrained parameters and their relative
(in)dependence. This analysis confirms what the confidence limits suggest:
the best-constrained parameters are the orbital frequency $\nu_\phi$, the hot
spot lifetime $T_l$, the arc length $\Delta \phi$, and the background flux
$I_B$. For the type A QPOs, we find $\nu_\phi$ and
$T_l$ to be independent, while the arc length and background flux are
strongly correlated, so that $\Delta \phi/I_B$ is positive and roughly
constant within our quoted confidence region. This is because, for
shorter arcs with fixed emissivity, increasing the arc length will
increase the amplitude of the light curve modulation, requiring a
larger background flux to give the same QPO amplitude. For the type B
QPOs on the other hand, a longer arc length does not significantly
amplify the modulation, since in the limit $\Delta \phi \to
360^\circ$, the light curve would remain constant, and thus the
parameters $\Delta \phi$ and $I_B$ are relatively independent. For
both type A and type B
QPOs, we find that the resonance width and the coronal scattering
length are independent, yet not very well constrained.

The resulting amplitudes and widths of the major QPO peaks are shown
in Table \ref{tablethree}, along with $1\sigma$ confidence limits. These
amplitudes are given by the analytic model so that
the total rms in the peak centered at $\nu_j$ is 
\begin{equation}\label{rms_j}
{\rm rms}_j = \sqrt{2}\frac{A_j'}{A_0'},
\end{equation}
where $A_0'$ is the mean amplitude of the light curve (including the
background $I_B$) and $A_j'$ are the original Fourier amplitudes $A_j$
given by the ray-tracing code, appropriately scaled according to
equations (\ref{poisson_PDS}) and (\ref{scatter_PDS}). This is more
instructive than measuring the rms directly from
$Q(\nu)$, which includes background power and instrumental effects
uncorrelated to the actual QPO peaks. 

In \citet{schni04a}, the hot spot light curve was added to a steady-state
disk with emissivity that scales as $r^{-2}$, which provides an
estimate of the size and overbrightness of the hot spots required to
produce a given (rms/mean) amplitude in the light curve. Considering that
most high frequency QPOs are observed with the greatest significance
in the 6-30 keV energy band during the steep power-law
spectral state \citep{mccli04}, it seems rather unlikely that the
background flux is coming from a thermal, optically thick disk. Even
if the flux is originally produced by such a disk, it clearly 
undergoes significant scattering in a hot corona to give the high
temperature power law observed in the photon energy spectrum.

In the context of the
model presented here, we can only calculate the fraction of the total flux
that is coming from the hot spots, determined by fitting to the QPO
data, without presuming an actual model for the background
emission. For XTE J1550-564, we find that the type A hot spot/arcs
contribute $8.5\%$ of the flux in the 6-30 keV band, while the type B
arcs must contribute significantly more $(38\%)$ to give a comparable
amplitude. This is due to the longer arc length described above: in
the limit of an azimuthally symmetric ring, even infinite brightness
would produce no variability.

\section{DISCUSSION AND CONCLUSIONS}

In the context of a geodesic hot spot model, we have developed a few
simple analytic methods to interpret the amplitudes and widths of QPO
peaks in accreting black holes. The model combines
three-dimensional ray tracing calculations in full general relativity
with analytic results of basic convolution theory, which are in turn
confirmed by simulating the observed light curves of multiple hot
spots. Given the Fourier amplitudes of a single hot spot, we have
derived a simple formula for the complete QPO power spectrum made
up of Lorentzian peaks of varying amplitudes and widths. This
power spectrum can then be fit to observed QPO data and used to
constrain parameters of the hot spot model, and possibly measure the
black hole mass and spin.

For XTE J1550-564, the locations of the HFQPO peaks are well
constrained, in turn constraining the spin parameter $a/M$ when
combined with radial velocity measurements of the black 
hole mass. Based on the presumption that the 3:2 frequency
ratio is indeed caused by closed orbits with coordinate frequencies in a
3:1 ratio, an observed mass of $M=10.5\pm 1.0 M_\odot$ and orbital
frequency $\nu_\phi=276\pm5$ Hz would predict a spin of $a/M=0.5\pm
0.1$ \citep{orosz02,remil02}. If reliable, this coordinate frequency
method would give one of the best estimates yet for a black hole spin.

The amplitudes of the QPO peaks can be used to infer the arc
length of the sheared hot spot and the relative flux contributions
from the hot spot and the background disk/corona. The longer arcs seen
in type B QPOs are also consistent with the broader peaks: if the hot
spots are continually formed and destroyed along special closed
orbits, as the emission region gets stretched into a ring, it is more
likely to be dissolved or disrupted, giving a shorter characteristic
lifetime $T_l$ and thus broader peaks.

Unfortunately, the quality of the QPO data is not sufficiently
high to confirm or rule out the present hot spot model, leaving a
number of questions unanswered. By fitting only two or three peaks, we are
not able to tightly constrain all the model parameters, particularly the
scattering length scale and the resonance width, both of which are
most sensitive to the higher frequency harmonics. Since the high
temperature electrons in the corona tend to transfer energy into the
scattered photons, measuring the energy spectra of the different QPO
peaks would also prove extremely valuable in understanding the
emission and scattering mechanisms. This has been done to some degree
with the lower frequency region of the power spectra from black holes
and neutron stars \citep{ford99}, and may even be observable above $\sim
100$ Hz with current RXTE capabilities. For this analysis to be most
effective, a more accurate model for the electron
scattering will certainly be necessary.

Some of the power spectrum features discussed in this paper are unique
to the geodesic hot spot model, while others could be applied to more
general QPO models. Clearly the harmonic amplitudes $A_j$ given by the
ray-tracing calculation are dependent on the hot spot model, as is the
broadening from a finite resonance width, yet both could be
generalized and applied to virtually any perturbed disk
model. Similarly, the random phase broadening and
the damping of higher harmonics due to photon scattering will be important
effects for any emission mechanism that produces periodic light
curves from black holes. If the next generation X-ray timing mission
could produce power
spectra comparable to the phenomenal cosmic microwave background (CMB)
results of recent years, we believe that many of these issues could be
resolved. Energy resolution and polarization would similarly provide
extremely valuable information about the source of the QPOs. As with
the CMB, each successive peak of the power spectrum would help to pin
down another parameter until the  model becomes \textit{predictive}
instead of descriptive, or is ruled out all together.

In the immediate future, however, there is much more to be done with
the RXTE data that already exists. Important additional insight might
be gained from new analysis of the X-ray light curves in the time
domain, recovering some of the phase information lost when the power
spectrum is computed in frequency space. There is also an important
message in the relationship between the photon energy spectra and the
QPO power spectra as well as the connection between the low frequency
and high frequency QPOs. Why should the HFQPOs appear in certain spectral
states and not others? The answer to these questions may lie in new
models of the accretion disk and specifically the radiation physics
relating the thermal and power-law emission, as well as broad
fluorescent lines like Fe K$\alpha$. The fact that the HFQPOs are seen
most clearly in the 6-30 keV energy range suggests that standard
models of thin, thermal accretion disks are not adequate for this
problem. This emphasizes the essential role of radiation transport,
particularly through the corona, in any physical model for black hole
QPOs.

\vspace{1cm}
We thank Ron Remillard for many helpful discussions and providing the
QPO data for XTE J1550-564. This work was supported by NASA
grant NAG5-13306.

\newpage

\begin{table}[tp]
\caption{\label{tableone} Widths of QPO peaks around coordinate frequency
modes $n\nu_\phi \pm \nu_r$, due to a radial distribution of hot spots with
$\Delta r = 0.05M$. For relatively narrow resonance regions, the QPO
peak widths are linearly proportional to $\Delta r$. The basic black
hole and hot spot model parameters are the same as in Figures
\ref{plotone} and \ref{plotthree}.}
\begin{center}
\begin{tabular}{lcc}
Mode & Frequency (Hz) & FWHM (Hz) \\
\hline
$\nu_r$ & 95 & 3.6 \\
$\nu_\phi-\nu_r$ & 190 & 12.2 \\
$\nu_\phi$ & 285 & 8.4 \\
$\nu_\phi+\nu_r$ & 380 & 4.8 \\
$2\nu_\phi-\nu_r$ & 475 & 20.6 \\
$2\nu_\phi$ & 570 & 16.8 \\
$2\nu_\phi+\nu_r$ & 665 & 13.2 \\
\end{tabular}
\end{center}
\end{table}

\begin{table}[tp]
\caption{\label{tabletwo} Best-fit parameters of the hot spot model for type
A and type B QPOs from XTE J1550-564. $(1\sigma)$ confidences are shown in
parentheses.}
\begin{center}
\begin{tabular}{lccc}
  Parameter & & Type A & Type B \\
  \hline
  orbital frequency $\nu_\phi$ (Hz) & & 280.1(2.4) & 270.5(12)\\
  lifetime $T_l$ (ms) & & 10(2.0) & 5(1.5) \\
  \hspace{1.6cm} (orbits) & & 2.8(0.55) & 1.4(0.4) \\
  resonance width $\Delta r$ ($M$) & & 0.02(0.05) & 0.025(0.12) \\
  scattering length $\lambda$ ($M$)& & 5(10) & 10(20) \\
  arc length $\Delta\phi$ ($^\circ$) & & 155(30) & 285(20) \\
  flux ratio $\frac{I_{\rm{hot spot}}}{I_B+I_{\rm{hot spot}}}$ & &
  0.085(0.025) & 0.38(0.05)   
\end{tabular}
\end{center}
\end{table}

\begin{table}[bp]
\caption{\label{tablethree} Amplitudes and widths of type A and type B QPO
peaks from XTE J1550-564, as determined by the best fit parameters
listed in Table \ref{tabletwo} and equation (\ref{rms_j}). $(1\sigma)$
confidences are shown in parentheses.}
\begin{center}
\begin{tabular}{lclclclcl}
  & & &A& & & &B& \\
  Mode & & rms & & FWHM & & rms & & FWHM \\
  & & (\%) & & (Hz) & & (\%) & & (Hz) \\
  \hline
  $\nu_r$ & & 0.57(0.15) & & 33.1(6.2)& &2.03(0.21)& & 63.6(16.0) \\
  $\nu_\phi-\nu_r$ & & 1.62(0.26) & & 35.7(5.9)& &2.57(0.14)& &67.6(15.5)\\
  $\nu_\phi$ & & 3.35(0.17) & & 34.6(5.5)& &1.48(0.24)& &65.9(15.3)\\
  $\nu_\phi+\nu_r$ & & 0.75(0.19) & & 33.4(5.8)& &0.06(0.02)& &64.1(15.8)\\
\end{tabular}
\end{center}
\end{table}

\newpage

\end{document}